\documentclass[aps,twocolumn,prb,floatfix,tightenlines,superscriptaddress,longbibliography]{revtex4-1}
\usepackage{graphicx}
\usepackage{epstopdf}
\usepackage[english]{babel}
\usepackage{amsmath}
\usepackage{amssymb}
\usepackage{mathrsfs}
\usepackage{appendix}
\usepackage{bm}
\usepackage{float}
\usepackage{color}
\usepackage{longtable}
\usepackage{url}
\usepackage[usenames,dvipsnames]{xcolor}
\usepackage[colorlinks=true,linkcolor=Blue,urlcolor=Blue,citecolor=Blue]{hyperref}

\graphicspath{{./Figures}}

\begin{document}

\title{Topological quantum phase transitions retrieved through unsupervised machine learning}

\author{Yanming Che}
\email{yanmingche01@gmail.com}
\affiliation{Theoretical Quantum Physics Laboratory, RIKEN Cluster for Pioneering
Research, Wako-shi, Saitama 351-0198, Japan}

\author{Clemens Gneiting}
\affiliation{Theoretical Quantum Physics Laboratory, RIKEN Cluster for Pioneering
Research, Wako-shi, Saitama 351-0198, Japan}

\author{Tao Liu}
\affiliation{Theoretical Quantum Physics Laboratory, RIKEN Cluster for Pioneering
Research, Wako-shi, Saitama 351-0198, Japan}

\author{Franco Nori}
\email{fnori@riken.jp}
\affiliation{Theoretical Quantum Physics Laboratory, RIKEN Cluster for Pioneering
Research, Wako-shi, Saitama 351-0198, Japan}
\affiliation{Department of Physics, The University of Michigan, Ann Arbor, Michigan
48109-1040, USA}

\date{\today}

\begin{abstract}
The discovery of topological features of quantum states plays an important 
role in modern condensed matter physics and various artificial systems. 
Due to the absence of local order parameters, the detection of topological 
quantum phase transitions remains a challenge. Machine learning may 
provide effective methods for identifying topological features. 
In this work, we show that the unsupervised manifold learning can successfully 
retrieve topological quantum phase transitions in momentum and real spaces. Our 
results show that the Chebyshev distance between two data points sharpens 
the characteristic features of topological quantum phase transitions in 
momentum space, while the widely used Euclidean distance is in general 
suboptimal. Then a diffusion map or isometric map can be applied to 
implement the dimensionality reduction, and to learn about topological quantum 
phase transitions in an unsupervised manner. We demonstrate this method on the 
prototypical Su-Schrieffer–Heeger (SSH) model, the Qi-Wu-Zhang (QWZ) model, 
and the quenched SSH model in momentum space, and further provide implications 
and demonstrations for learning in real space, where the topological invariants 
could be unknown or hard to compute. The interpretable good performance of 
our approach shows the capability of manifold learning, when equipped with 
a suitable distance metric, in exploring topological quantum phase transitions. 
\end{abstract}

\maketitle

\section{Introduction}
Topological phases of matter have attracted tremendous 
attention in the past decade~\cite{KaneRMP2010,QiRMP2010,HaldanePRL2008,PollmannPRB2010,Mousavi2015,MalzardPRL2015,LeykamPRL2017,LiuPRB2017,YuRanPRL2017,Schindler2018,ZhongWangPRL2018,TaoLiuPRL2018,OzawaRMP2019,TaoLiuPRL2019,BliokhNC2019}. Conceptually, topological quantum phase transitions go beyond the conventional Landau paradigm, which needs local order parameters to distinguish different phases. Instead, topological quantum phases are usually characterized by topological quantum numbers, which reflect global properties of the state manifold defined on a compact Brillouin zone. Quantum states belonging to the same topological sector can be continuously deformed to each other without closing the bulk energy gap, and are thus called homotopic. When a topological quantum phase transition occurs, according to the bulk-boundary correspondence, the band gap closes at the critical point. Therefore, a topological quantum phase transition is usually accompanied by a discontinuous change of the state configuration, such as the sign change of the mass term in the Hamiltonian, or band inversions in topological insulators~\cite{KaneRMP2010, QiRMP2010}. 

Due to the absence of local order parameters, the detection of topological quantum phase transitions remains a challenge. For a given model Hamiltonian in momentum space featuring a set of parameters, it is usually not obvious whether it exhibits topological quantum phase transitions when sweeping the model parameters. Recently, machine learning has been introduced to quantum physics, including seeking quantum-enhanced learning algorithms~\cite{Schuld2014,YuTing2017,Biamonte2017QML,Schuld2017,Dunjko2018}, and detecting quantum phase transitions with machine learning in many-body physics~\cite{CarleoRMPML,LeiWangPRB2016,CostaPRB2017,TroyerScience2017, YeHuaNatPhys2017,CarrasquillaNatPhys2017,WetzelPRE2017Unsupervised,HuPRE2017Discovering,PontePRB2017Kernel,WangPRB2017Machine,PhysRevX.7.031038,Broecker2017,YeHuaPRL2018,VenderleyPRL2018,XinWanPRB2018,KelvinPRB2018,FrankPRE2019,DurrPRB2019,ZhangPRE2019Machine,HuembeliPRB2019,PhysRevB.99.121104,RemNatPhys2019,ming2019quantum,MehtaPhysRep2019,PhysRevB.99.041108,PhysRevB.99.104410,Giannetti2019,PhysRevB.100.045129,Jadrich2018I,Jadrich2018II,Ohtsuki2020,arxiv:1903.03506,Munoz_Bauza_2020,arxiv:1910.13453} and topological systems~\cite{Klintenberg2014,Ohtsuki2016,ZhangPRL2017Quantum,HuiZhaiPRL2018,BeachPRB2018Machine,HuiZhaiPRB2018,HuembeliPRB2018,YoshiokaPRB2018,Pilozzi2018,Rodriguez-NievaNatPhys2019,BalabanovPRR2020,GreplovaPredictive2020,ZhangInterpreting2020}. In particular, deep learning has been employed~\cite{HuiZhaiPRL2018,ZhangPRL2017Quantum} to train a neural network to recognize different topological phases in a supervised manner. 

However, prior knowledge and labeled training examples are not always easily accessible in practice. In that respect, the unsupervised learning, without pre-training, provides a more promising learning framework to discover topological patterns. For instance, neural networks with autoencoders~\cite{arxiv:1908.00281} and predictive models~\cite{GreplovaPredictive2020} were used to learn topological features without explicit supervisions. 
Compared with deep learning, manifold learning methods, such as Isomaps~\cite{TenenbaumScience2000} and diffusion maps~\cite{coifman2005geometric,coifman2006diffusion}, usually require lower computational costs. The unsupervised learning based on diffusion maps has been successfully applied for the identification of topological clusters~\cite{Rodriguez-NievaNatPhys2019}. The application of this method leaves the freedom to choose the distance metric in the original feature space. The Euclidean distance (ED), which has been used in most cases, appears as a natural choice. 

In this work, we propose to use the better optimized Chebyshev distance (CD) to construct the graph-structured data set, and 
to use unsupervised manifold learning to retrieve topological quantum phase transitions in momentum space, which can be further extended, in an interpretable manner, to learning in real space. The use of the CD was inspired by the fidelity-susceptibility
indicator~\cite{MaPRB2010} for topological quantum phase transitions, as well as the non-Euclidean structure of the data set. 

We demonstrate this method on the prototypical Su-Schrieffer–Heeger (SSH) model~\cite{SSH1980,Asbth2016}, the two-dimensional ($2$D) Qi-Wu-Zhang (QWZ) model~\cite{QWZ}, and the quenched SSH model~\cite{GongPRL2018} in momentum space, and further provide implications and demonstrations for learning in real space, where the topological invariants could be unknown or hard to compute. The interpretable good performance of our approach highlights the capability and promising performance of unsupervised machine learning of topological quantum phase transitions, without prior analysis of the Hamiltonian.

\section{Manifold learning and distance metric}
Manifold learning is used when linear unsupervised learning models, like the principal component analysis, fail to uncover nonlinear structures in data sets~\cite{TenenbaumScience2000,RoweisScience2000}. A key characteristic of manifold learning methods such as Isomap~\cite{TenenbaumScience2000} is that the ED is not suitable to reflect the intrinsic connectivity and similarity between data points. Adapted distance metrics such as the manifold geodesic distance, approximated from the shortest distance on the neighborhood graph, should then be used to characterize the similarities. The following step is to embed the data points into a meaningful low-dimensional Euclidean space based on this distance or similarity matrix, where a conventional clustering method (e.g., $k$-means) can be used to detect clusters in the data set. 

When it comes to the unsupervised learning of topological quantum phase transitions for states in momentum space, one comes across a similar problem: \emph{the ED in general cannot successfully retrieve topological clusters in the data set}. 
A homotopic distance metric (see e.g., Ref.~\onlinecite{arxiv:1810.12591}) is instead needed to adequately capture the structure of the data set. While the generic numerical evaluation of the homotopic distance between two input vectors is difficult at the current stage, some approximative metrics may be used. 

Motivated by the observation that topological transitions usually are accompanied by \emph{sign changes or band inversions}, here we investigate the use of the $\mathbb{L}^{\infty}$-norm induced CD to (approximately) measure topological similarities. Across topological transitions, quantum states defined over the compact Brillouin zone (BZ) take sharp changes at certain symmetric points~\cite{SSH_transition_note} in the BZ, while state vectors belonging to the same topological phase vary smoothly. The CD \emph{highlights}~\cite{EDNote} these features of topological quantum phase transitions, facilitating the successful retrieval of the critical lines. 

To see this more clearly, we analyze the performance of the distance metric in clustering with the similarity (or kernel) matrix  
\begin{equation}
\label{eq:K_matrix}
{\cal{K}}^p_{ij} = \mathrm{exp}\left(- \ \frac{\lVert\mathbf{x}_i - \mathbf{x}_j \rVert^2_{\mathbb{L}^p}}{\epsilon}\right),
\end{equation}
where $\epsilon$ is the resolution parameter of the kernel, $\textbf{x}_i$ is the $i$th input data vector and $i, j = 0, 1, 2, ..., M-1$, with $M$ the size of the data set. The more similar (or closer) two data points are, the larger value their corresponding kernel matrix element will take. The $\mathbb{L}^p$ norm of a vector $\textbf{z} = \left( z_1, \cdot \cdot \cdot, z_{d_z} \right)$ is given by 
\begin{equation}
\lVert \textbf{z} \rVert_{\mathbb{L}^p}=\left(\sum_{k=1}^{d_z} \left | z_k \right |^p\right)^{1/p},
\end{equation}
where $p=2$ and $p=\infty$ give the ED and the CD, respectively. The $\mathbb{L}^p$ norms with $p=1, 2, \infty$ are widely used due to their explicit geometric and physical meanings. An equivalent but more useful expression for the $\mathbb{L}^{\infty}$ norm is 
\begin{equation}
\lVert\textbf{z}\rVert_{\mathbb{L}^{\infty}}=\mathop{\mathrm{max}}\limits_k  \left | z_k \right |.  
\end{equation}
\begin{figure}[t]
\centerline{\includegraphics[height=5in,width=3.3in,clip]{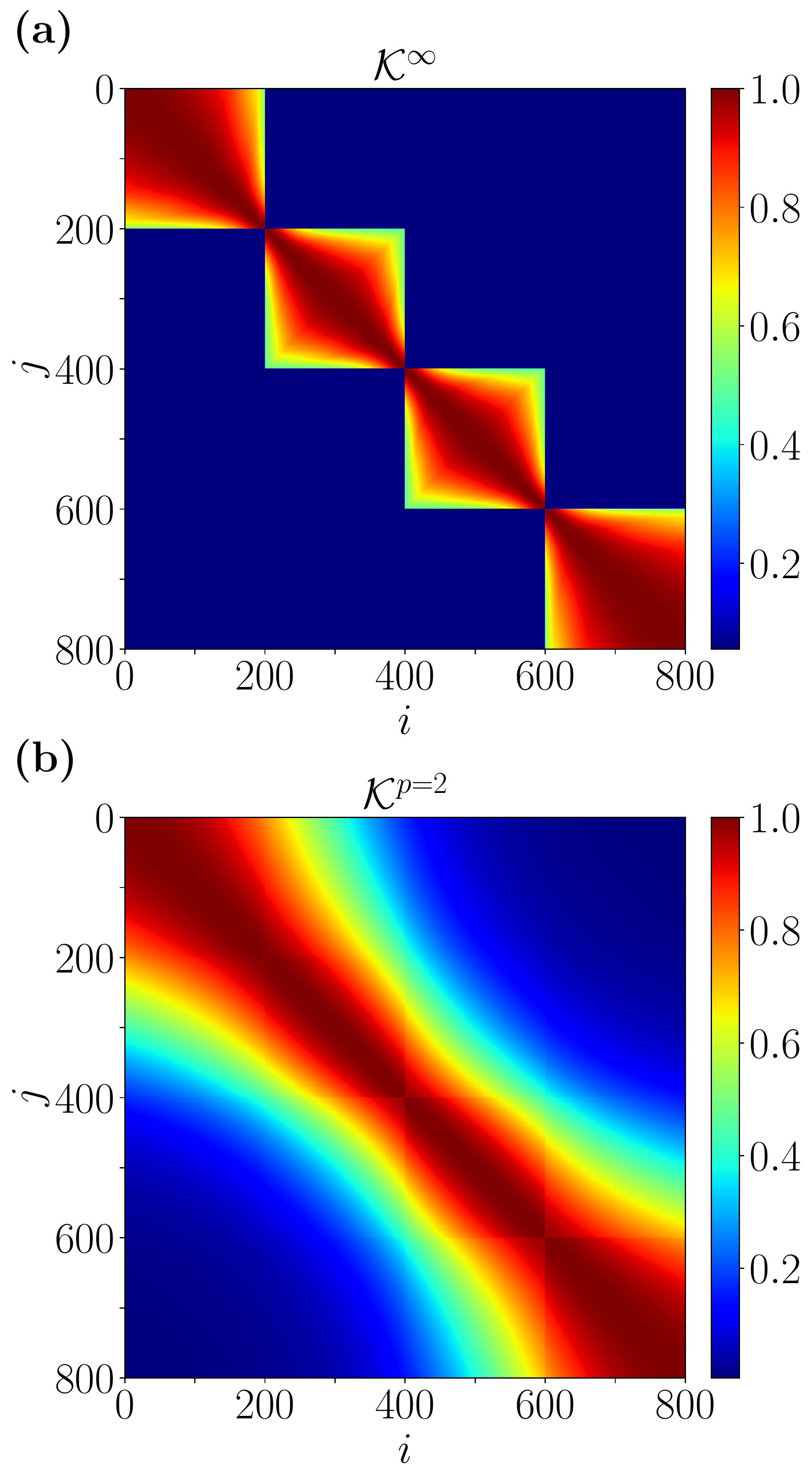}}
\caption{(Color online) Similarity matrix ${\cal{K}}^p$ in Eq.~(\ref{eq:K_matrix}) built from the $\mathbb{L}^p$-norm distance metric: Compared are the (a) Chebyshev distance (CD) with $p=\infty$ and (b) the Euclidean distance (ED) with $p=2$. $M=800$ input feature vectors (i.e., reshaped unit Bloch vectors; see the main text) are uniformly sampled in an \emph{ordered} manner, from the $2$D Qi-Wu-Zhang model in Eq.~(\ref{eq:d_component}), where the hopping energy is taken to be $b=0.2$ and the chemical potential $\mu$ varies from $\mu=-4b$ to $\mu=12b$, and the Brillouin zone is sliced into $32 \times 32$ patches. The resolution parameters are $\epsilon = 0.345$ in (a) and $\epsilon = 1.28 \times 10^{-4}$ in (b), which are obtained by minimizing the respective mean squared errors with respect to the ideal similarity matrix, where matrix elements for intra-cluster data points equal one and equal zero for the others. It can be seen that, compared to the ED, the CD sharpens the cluster boundaries as well as the characteristic feature of topological quantum phase transitions.}
\label{fig:K_matrix}
\end{figure}

\begin{figure}[t]
\centerline{\includegraphics[height=4.0in,width=3.5in, clip]{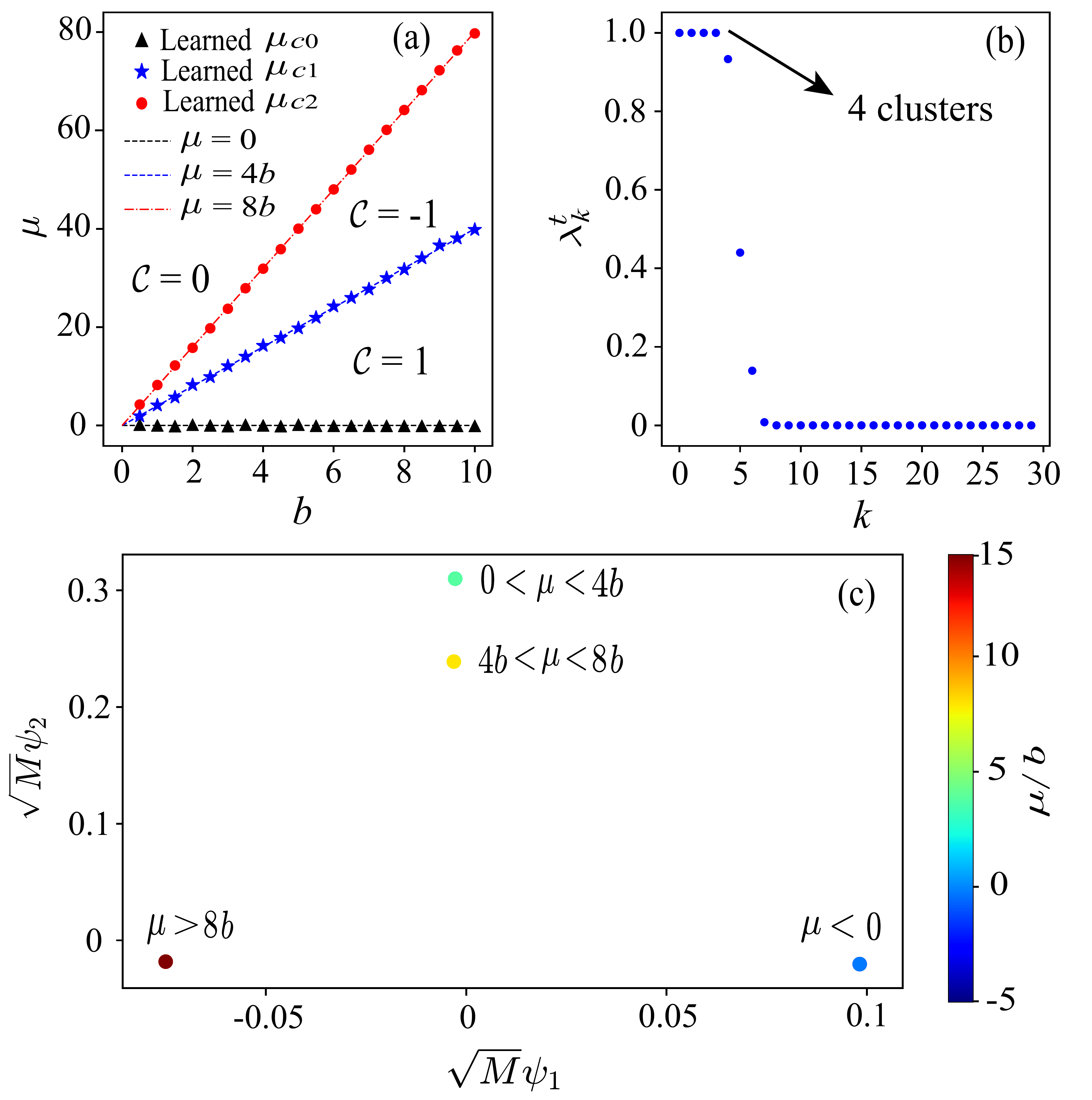}}
\caption{(a) Topological quantum phase transitions of the Qi-Wu-Zhang model retrieved from the diffusion map. The critical lines are automatically learned by scanning the hopping energy $b$ and for each value of $b$, the input data set consists of uniformly sampled feature vectors from $\mu=-5b$ to $\mu=15b$ (see the main text for details)~\cite{ParameterRangeNote}. The critical values of the chemical potential $\mu$ are identified by $k$-means in the diffusion space from the boundaries of $\mu$ parameters in each cluster. The first topologically trivial phase with $\mu < 0$ and ${\cal{C}}=0$ is not shown. (b) First $30$ eigenvalues of the diffusion matrix $P^t$. The degeneracy of the largest $\lambda_k^t$ indicates that there are $4$ disconnected clusters in total. (c) Images of the data set in the low-dimensional diffusion space. The colors encode the values of $\mu/b$, as indicated in the colorbar in the right-hand side of (c). The clustering is very effective such that the size of each cluster is far smaller than their inter distance. There are in total four clusters in (c), where each sector contains several hundreds of dots, with the values of $\mu/b$ (and also colors) gradually varying, and dots with lighter colors are covered by those with darker ones in each cluster. (b) and (c) are plotted with a representative value of $b=1$. The hyperparameters used are $M=1000$, $N=32$, $\epsilon=0.03$, and $t=500$. In contrast, the use of the Euclidean distance does not deliver the correct clustering, independent of the choice of $\epsilon$ (see Appendix~\ref{sec:append_QWZ_ED}).}
\label{fig:QWZ_diff_map}
\end{figure}

For instance, we show the similarity matrix ${\cal{K}}^p$ built upon the CD in Fig.~\ref{fig:K_matrix}(a) and upon the ED in Fig.~\ref{fig:K_matrix}(b) for the $2$D QWZ model~\cite{QWZ} [see Eq.~(\ref{eq:d_component})]. The data points are uniformly sampled unit Bloch vectors in an ordered way such that one can see four equally partitioned clusters. The resolution parameters are optimized with respect to the ideal similarity matrix (see the caption of Fig.~\ref{fig:K_matrix} for details). Compared to the ED displayed in Fig.~\ref{fig:K_matrix}(b), the similarity matrix from the CD in Fig.~\ref{fig:K_matrix}(a) clearly shows four nearly ideal clusters with good (poor) intra- (inter-) cluster connectivity, which correspond to four distinct sectors in the topological phase diagram in Fig.~\ref{fig:QWZ_diff_map}(a). The ED leads to connected clusters, and we find that, even for smaller values of $\epsilon$, the first two sectors with the ED are connected and thus cannot be correctly clustered by the diffusion-map algorithm (see Appendix~\ref{sec:append_QWZ_ED} for details). 

Note that when the algorithm is fed with unit Bloch vectors, the distances need to be normalized, which leads to 
\begin{subequations}
\begin{equation}
{\cal{K}}_{ij}^{\infty}= \mathrm{exp}\left(- \ \frac{\lVert \mathbf{x}_i - \mathbf{x}_j \rVert^2_{\mathbb{L}^{\infty}}}{4\epsilon}\right),
\end{equation} 
and 
\begin{equation}
{\cal{K}}_{ij}^{p=2}= \mathrm{exp}\left(- \ \frac{\lVert \mathbf{x}_i - \mathbf{x}_j \rVert^2_{\mathbb{L}^2}}{4\epsilon\left(N+1\right)^{2D}}\right),
\end{equation}
\end{subequations}
where the BZ is sliced into $N^D$ patches in the $D$-dimensional case. It can be shown straightforwardly that 
${\cal{K}}_{ij}^{p=2} \ge {\cal{K}}_{ij}^{\infty}$ with the same value of $\epsilon$, and the ED overestimates the inter-cluster connections, as compared to the CD.

\section{Diffusion map and dimensionality reduction}
With the CD as a topologically viable distance measure, we are now ready to seek an appropriate approach for the dimensionality reduction, to learn the topological clusters. Among many manifold learning methods~\cite{TenenbaumScience2000,RoweisScience2000,Maaten2008t-SNE,belkin2002laplacian,belkin2003laplacian}, diffusion maps~\cite{coifman2005geometric,coifman2006diffusion,nadler2006diffusion,farbman2010diffusion,haghverdi2015diffusion,Rodriguez-NievaNatPhys2019} can successfully discover the connected components in a data manifold if a viable distance measure is used, visualized by the similarity matrix in Fig.~\ref{fig:K_matrix}. In the framework of diffusion maps, a Markovian random walk is launched within the data set, where the transition probability between two data points is given by the normalized similarity matrix, 
\begin{equation}
P_{ij}=\frac{ {\cal{K}}^p_{ij} } { \sum_j {\cal{K}}^p_{ij} }.
\end{equation}
After $t$ steps of random walk, the connectivity between two points $\textbf{x}_i$ and $\textbf{x}_j$ is characterized by the diffusion distance~\cite{nadler2006diffusion}
\begin{equation}
D_t^2\left(\textbf{x}_i, \textbf{x}_j\right) = \sum_k \frac{ [\left(P^t\right)_{ik}-\left(P^t\right)_{jk}]^2 } { \phi_0 \left(\textbf{x}_k\right) },
\end{equation}
where $\phi_0 \left(\textbf{x}_k\right)$ is the first left eigenvector of $P$. The map to the Euclidean space $Y$, which best preserves the connectivity of the data set, is determined by minimizing the cost function~\cite{Maaten2008t-SNE} 
\begin{equation}
\label{eq:cost_function}
C = \sum_{ij} \left[ D_t\left(\mathbf{x}_i, \mathbf{x}_j\right) - d_Y \left(\mathbf{y}_i, \mathbf{y}_j\right) \right] ^2,
\end{equation}
where $d_Y \left(\textbf{y}_i, \textbf{y}_j\right)$ is the $Y$-space Euclidean distance between the images $\textbf{y}_i$ and $\textbf{y}_j$ of two data points. The solution of minimizing the above cost function is given by~\cite{nadler2006diffusion} 
\begin{equation}
\textbf{y}_i=\left[\lambda_0^t \psi_0\left(\textbf{x}_i\right), \lambda_1^t \psi_1\left(\textbf{x}_i\right), ..., \lambda_{M-1}^t\psi_{M-1}\left(\textbf{x}_i\right)\right],
\end{equation}
where $\lambda_k$ and $\psi_k\left(\textbf{x}_i\right)$ are the $k$th eigenvalue and the right eigenvector's $i$th component of the $P^t$ matrix, respectively. 

Note that the eigenvalues of the probability matrix $P$ satisfy $0 \le \lambda_k \le 1 \ \forall k$. Hence when 
the number of diffusion steps $t$ is large, the nontrivial vector components of $\textbf{y}_i$ are given by those with 
$\lambda_k \approx 1$, and the degree of (near) degeneracy of these eigenvalues equals to the number of connected components in the similarity matrix~\cite{Rodriguez-NievaNatPhys2019}. The first eigenvector is $\psi_0 \propto 1/M = \mathrm{const}$, so data images are not fully dispersed in this dimension. Statistics and clustering methods such as $k$-means will be then used in the $Y$ space to identify the samples in each cluster, and the critical lines are decided \emph{automatically} by the corresponding cluster boundaries in the parameter space.
\begin{figure}[t]
\centerline{\includegraphics[height=4.0in,width=3.5in,clip]{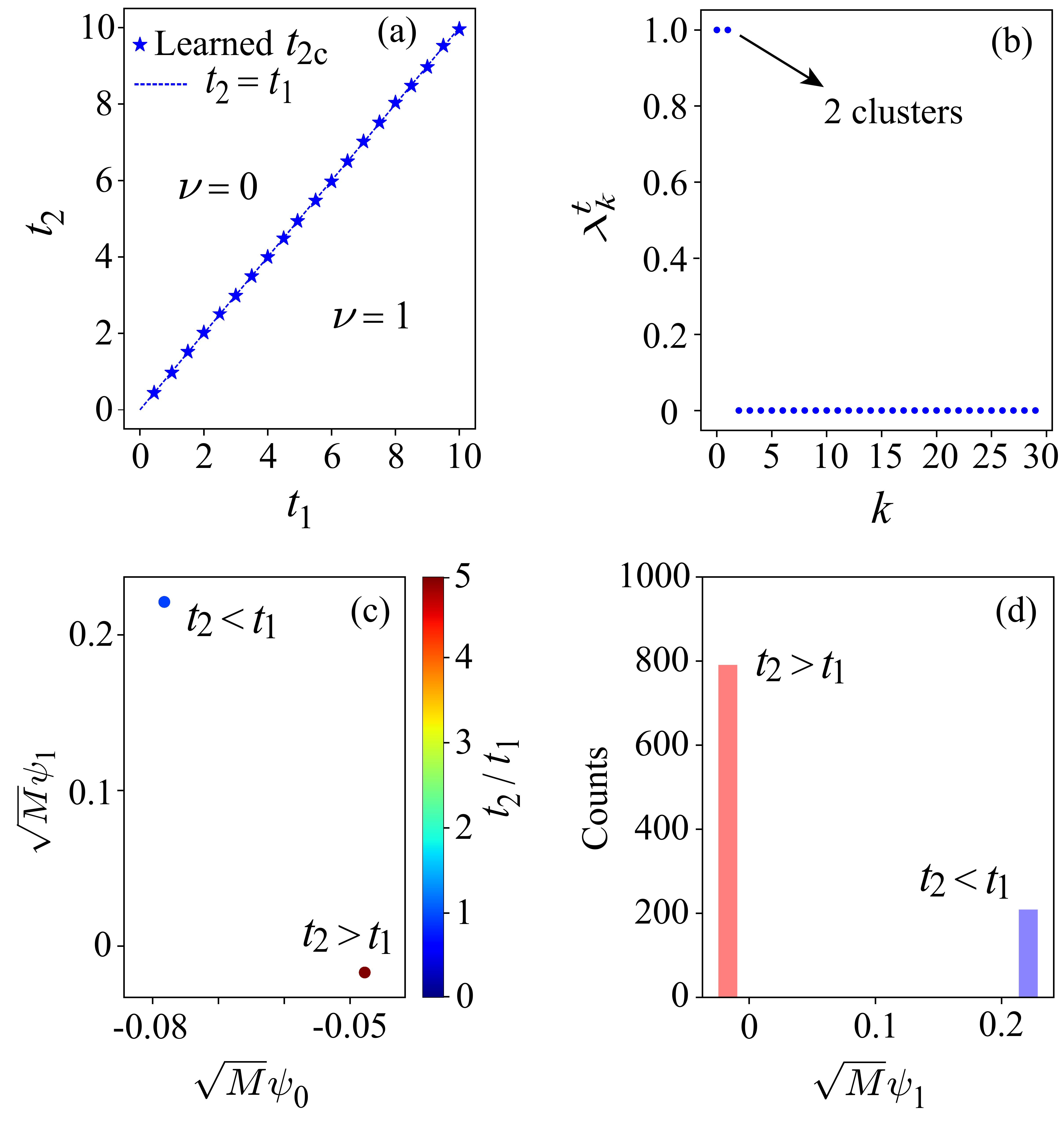}}
\caption{(a) Learned topological quantum phase transition of the SSH model. The critical line is automatically learned by scanning the hopping energy parameter $t_1$ and for each $t_1$, the input data set is composed of uniformly sampled feature vectors from $t_2 \in \left(0, 5t_1 \right]$ (see the main text for details). There are only two topological sectors detected, as illustrated by the degree of degeneracy of the largest eigenvalue $\lambda_k^t$ in (b). (c) Images of the data set in the embedded diffusion space. The colors encode the values of $t_2/t_1$, as indicated in the colorbar in the right-hand side of (c). The clustering is very effective such that the size of each cluster is far smaller than their inter distance. There are in total two clusters detected in (c), where each sector contains several hundreds of dots [see (d)], with the values of $t_2/t_1$ (and also colors) gradually varying, and the green and yellow dots are covered by the darker brown ones in the bottom right. The $k$-means clustering method is used in (c) to automatically retrieve the topological transition in (a). (b) and (c) are plotted with a representative value of $t_1=1$. Other parameters used are $M=1000$, $N=32$, $\epsilon=0.03$, and $t=500$.}
\label{fig:SSH}
\end{figure}

\section{Learning topological quantum phase transitions in two dimensions} 
First we consider the Qi-Wu-Zhang model~\cite{QWZ} in two dimensions. The Hamiltonian in momentum space is 
\begin{equation}
\label{eq:QWZ_H}
H\left(\mathbf{k}\right)=d_0\left(\mathbf{k}\right)+\mathbf{d}\left(\mathbf{k}\right) \cdot \bm{\sigma},
\end{equation}
with $\bm{\sigma}=\left(\sigma_x, \sigma_y, \sigma_z\right)$ the vector of Pauli matrices and $\textbf{d}\left(\textbf{k}\right)$ a three-dimensional vector with components
\begin{eqnarray}
\label{eq:d_component}
d_x&=&\sin\left(k_x\right), \ \ \ d_y=\sin\left(k_y\right), \nonumber \\ 
d_z&=&\mu-2b\left[2-\cos\left(k_x\right)-\cos\left(k_y\right)\right], 
\end{eqnarray}
where $\mu$ is the chemical potential and $b$ is the hopping energy, and we have taken a unit lattice constant. The $2$D BZ is given by $\left[-\pi, \pi\right] \times \left[-\pi, \pi\right]$. We can further assume $d_0\left(\textbf{k}\right)=0$, because it trivially contributes to the topology. The normalized unit vector $\hat{\textbf{d}}\left(\textbf{k}\right)=\textbf{d}\left(\textbf{k}\right)/\left| \textbf{d}\left(\textbf{k}\right)\right|$ defines a mapping from the compact $2$D BZ (i.e., $\mathbb{T}^2$) to the unit sphere $\mathbb{S}^2$. The topological Chern number
\begin{equation}
\label{eq:2D_Chern_number}
{\cal{C}}_{2D} = \frac{1}{4\pi} \int_{BZ} \mathrm{d}k_x \mathrm{d}k_y \ \hat{\mathbf{d}} \cdot \left(\partial_{k_x}\hat{\mathbf{d}} \times \partial_{k_y}\hat{\mathbf{d}}\right)
\end{equation}
measures how many times the mapping wraps over the unit sphere. 

Shown in Fig.~\ref{fig:QWZ_diff_map}(a) are the topological critical lines \emph{automatically} learned in an unsupervised manner by scanning through the hopping energy $b$. For fixed $b$, the data set consists of $M$ uniformly sampled $\hat{\textbf{d}}$ vectors from $\mu=-5b$ to $\mu=15b$, and $\hat{\textbf{d}}\left(\textbf{k}\right)$ is vectorized to a high-dimensional feature vector in $\mathbb{R}^{3\left(N+1\right)^2}$, where we have discretized the BZ into $N \times N$ patches. The learned black triangles, red dots and blue stars precisely trace three critical lines that divide the parameter $(\mu, b)$ space into four topological sectors. 
\begin{figure*}[t]
\centerline{\includegraphics[height=2in,width=7in,clip]{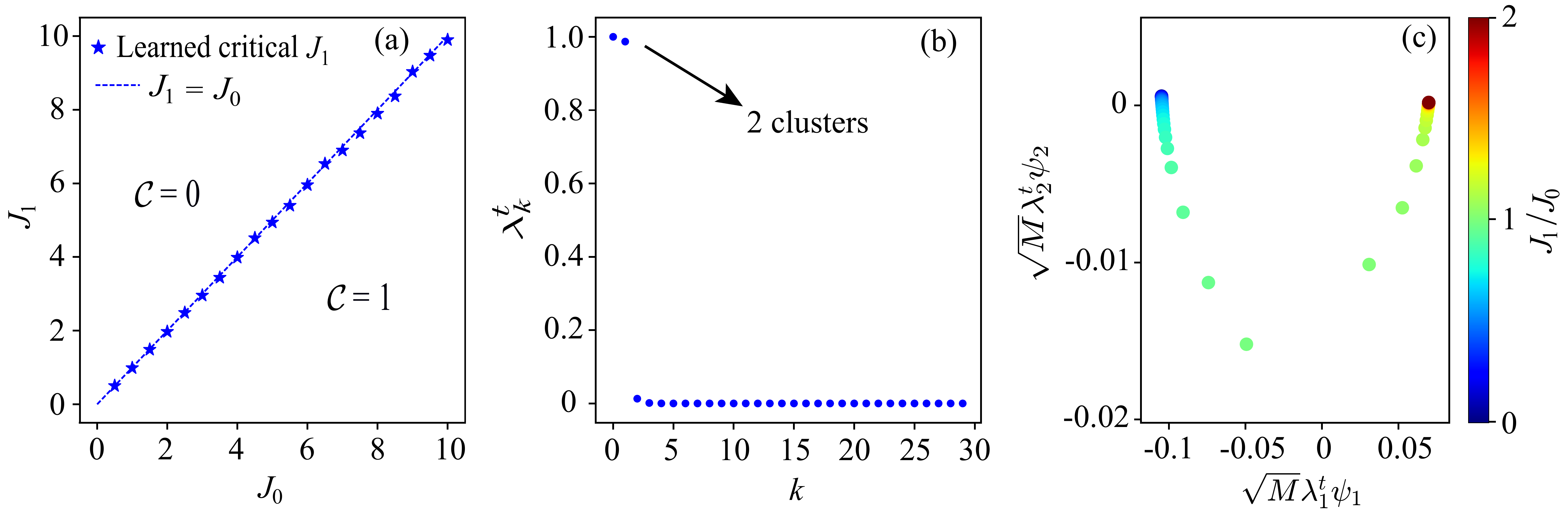}}
\caption{(a) Topological quantum phase transition of a quenched SSH model, retrieved from the diffusion map. Only two topological sectors are detected. The transition line is automatically learned by varying the parameter $J_0$, and for each fixed value of $J_0$ our input data set consists of sampled feature vectors (i.e., reshaped unit Bloch vectors) from $J_1 \in \left(0, 2J_0 \right]$. (b) shows the first $30$ eigenvalues $\lambda_k^t$ with $t=100$. The number of (near) degenerate largest eigenvalues indicate that there are two topological sectors. (c) Images of the data set in the embedded low-dimensional flat space. The $k$-means clustering method is used to automatically identify each cluster and the corresponding parameter $J_1$ for each data point. The critical line in (a) is obtained from the identified $J_1$ boundaries in each cluster. The color code indicates different values of $J_1/J_0$. (b) and (c) are plotted with a representative value of $J_0=1$. Here we have used $M=1000$, $N=32$, $\epsilon=0.03$.}
\label{fig:quench_SSH}
\end{figure*}

It should be noted that \emph{further topological information} may be retrieved from Fig.~\ref{fig:K_matrix}(a), where the first and fourth, the second and third clusters are quasi-symmetric, respectively, indicating that the corresponding two sectors in the phase diagram may have related topology. 

This can be verified by calculating the Chern numbers within each phase, as shown in Fig.~\ref{fig:QWZ_diff_map}(a). The number of degenerate eigenvalues with $\lambda^t \approx 1$ in Fig.~\ref{fig:QWZ_diff_map}(b) determines the number of distinct topological sectors. Fig.~\ref{fig:QWZ_diff_map}(c) shows the data set embedded in the diffusion space. The size of each cluster is sufficiently small such that points belonging to the same sector almost collapse. Compared to supervised learning, here the learning is automatically achieved without prior training. In Appendix~\ref{sec:append_QWZ_Isomap}, we also provide the principal component analysis (PCA) and an Isomap learning, respectively, for the QWZ model, where in the latter, results from the CD and the ED are compared.

\section{Learning the Su-Schrieffer–Heeger model}
The SSH model describes electrons in an one-dimensional lattice~\cite{SSH1980,Asbth2016}, with $t_1$ and $t_2$ the staggered hopping energies. The Hamiltonian of the SSH model in momentum space reads 
\begin{equation}
H(k)=\left[t_1+t_2 \cos(k)\right] \sigma_x + t_2 \sin(k) \sigma_y.
\end{equation}

In the Bloch vector formulation as in the $2$D case, we have 
\begin{eqnarray}
d_x(k) &=& \left[t_1+t_2 \cos(k)\right], \nonumber \\ 
d_y(k) &=& t_2 \sin(k), \\
d_z(k) &=& 0.  \nonumber 
\end{eqnarray}
Note that the absence of the third component is related to the chiral symmetry of the SSH Hamiltonian. The unit vector $\hat{\textbf{d}}\left(k\right)=\textbf{d}\left(k\right)/\left| \textbf{d}\left(k\right)\right|$ defines a map $\hat{\textbf{d}}\left(k\right): S^1 \mapsto S^1$, where the topological winding number is~\cite{Asbth2016}
\begin{equation}
\label{eq:1D_winding_number}
\nu = \frac{1}{2\pi i} \int_{-\pi}^{\pi} \mathrm{d}k \ q^{-1}(k) \partial_k q(k), 
\end{equation}
where $q(k) =d_x(k)-id_y(k)$. The input data is given by the reshaped $\hat{\textbf{d}}(k)$ ($k \in \mathrm{BZ}$) vector in high-dimensional feature space $\mathbb{R}^{2\left(N+1\right)}$, after slicing the BZ into $N$ patches. For fixed $t_1$, the data set is obtained from uniformly sampled feature vectors from $t_2=0$ to $t_2 = 5t_1$. 

Shown in Fig.~\ref{fig:SSH}(a) is the learned phase diagram indicating that there are only two topologically distinct sectors (also indicated in Fig.~\ref{fig:SSH}(b) by the degeneracy of $\lambda^t$ with a large $t$). Figs.~\ref{fig:SSH}(c)-(d) show the distribution of data points in the low-dimensional diffusion space. The critical line in Fig.~\ref{fig:SSH}(a) is decided by fixing $t_1$ and identifying the $t_2$ parameters in each cluster in Fig.~\ref{fig:SSH}(c). The obtained topological quantum phase transition coincides with the theoretical prediction~\cite{Asbth2016}. 

\section{Learning the quenched SSH model} 
As a final demonstration, we provide the unsupervised learning of the dynamically quenched SSH model~\cite{GongPRL2018}. Here the hopping energies $\left(t_1, t_2\right)=\left(J_0, 0\right)$ in the SSH model experience a sudden change during the quench, which leads to the pre- and post-quenched Bloch vectors $\textbf{d}(k)=\left(J_0, 0, 0\right)$ and 
\begin{equation}
\textbf{d}^{\prime}(k)=\left[J_1+J_0\cos(k), J_0\sin(k), 0\right],
\end{equation}
respectively. 

With the K-theory classification of quench dynamics, the parent Bloch Hamiltonian reads $H\left(t, k\right)= \textbf{d}\left(t, k\right) \cdot \bm{\sigma}$, where the components of $\textbf{d}\left(t, k\right)$ are given by~\cite{GongPRL2018} 
\begin{eqnarray}
\label{eq:quench_d_component}
d_x(t, k)&=&-J_0 +2J_0\left[J_0\sin(k) \chi(t, k)\right]^2, \nonumber \\ 
d_y(t, k)&=&-2J_0^2\left[J_1+J_0\cos(k)\right]\sin(k) \chi(t, k)^2, \\
d_z(t, k)&=&-J_0^2\sin(k)\sin\left[2d^{\prime}(k) t\right]/d^{\prime}(k),  \nonumber 
\end{eqnarray}
where 
\begin{equation}
\chi(t, k)=\sin\left[d^{\prime}(k) t\right]/d^{\prime}(k),
\end{equation}
with $d^{\prime}(k)$ the norm of $\textbf{d}^{\prime}(k)$, and $(t, k) \in \left[0, \pi/d^{\prime}(k)\right] \times \left[0, \pi\right]$. 

The same learning protocol as above can be used to learn the phase transition in $\left(J_0, J_1\right)$ 
space, and shown in Fig.~\ref{fig:quench_SSH}(a) is the retrieved critical line (blue stars), which fits the theoretical result (dashed blue line) well. The dynamical topological numbers ${\cal{C}}$ in each sector are calculated with the method provided in Ref.~\onlinecite{GongPRL2018}. 

Figure~\ref{fig:quench_SSH}(b) shows that there are two topological sectors in the phase diagram (we have taken $t=100$). Figure~\ref{fig:quench_SSH}(c) displays the diffusion-space distribution of uniformly sampled $M=1000$ data points from $J_1=0$ to $J_1=2J_0$ with $J_0=1$. The $k$-means clustering method is used in this low-dimensional Euclidean space to learn the critical line automatically.

In the above examples, the data sets consist of normalized Bloch vectors. In Appendix~\ref{sec:append_wavefunction}, we also provide distance and similarity measures (i.e., the CD and the ED) for learning over \emph{wave functions} in momentum space.

\section{Implications for learning in real space}
We have presented manifold learning for topological quantum phase transitions of several prototypical models in momentum space.
While the ED may, in simple cases and for fine-tuned choices of $\epsilon$, deliver correct clustering, the CD is much more consistent and generic in its performance, as the $\mathbb{L}^{\infty}$-norm distance metric captures the characteristic features of topological quantum phase transitions for states in momentum space. 
\begin{figure}[t]
\centerline{\includegraphics[height=3in,width=3.5in,clip]{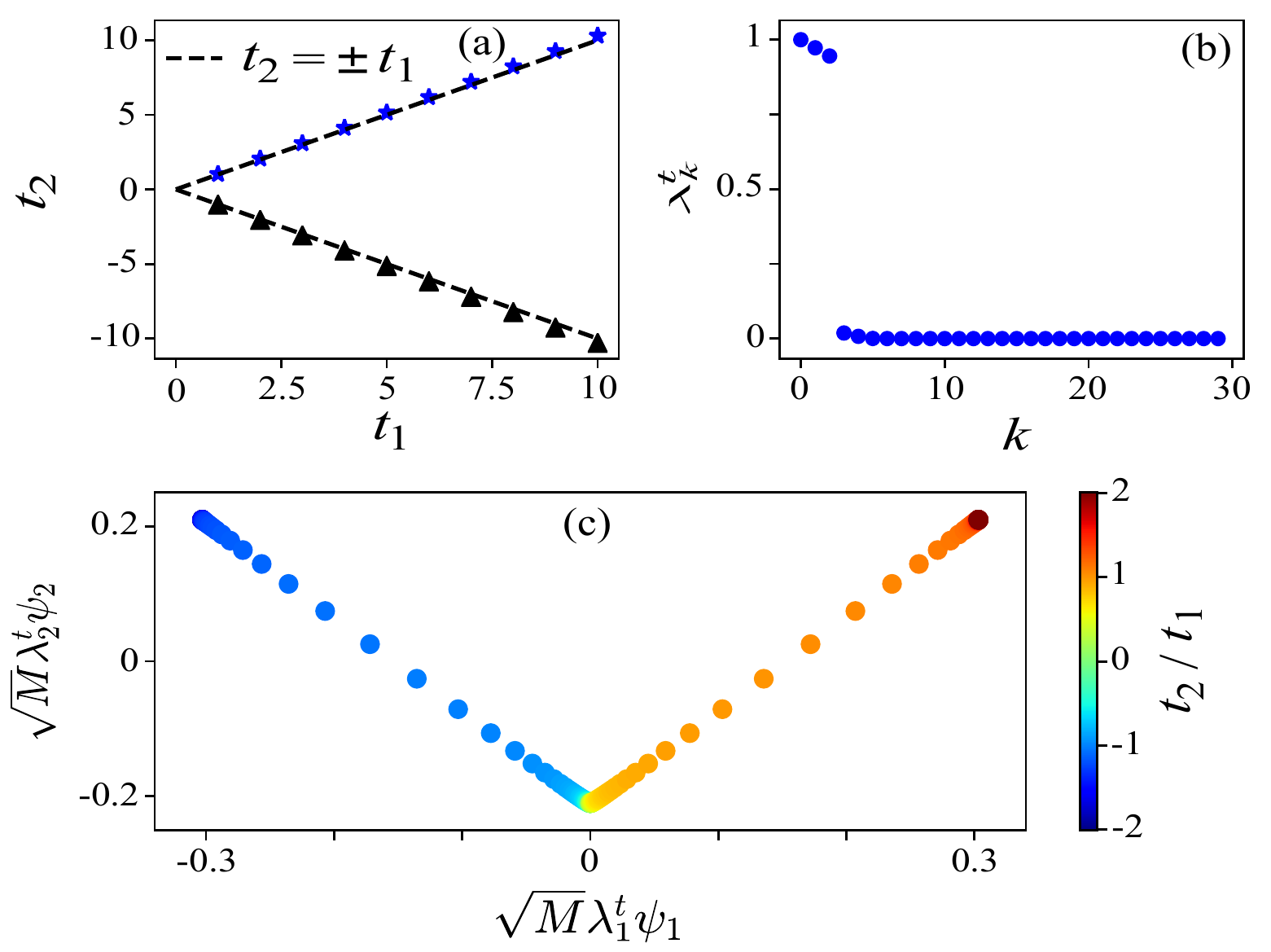}}
\caption{(a) Unsupervised learning of transition lines of the model in Eq.~(\ref{eq:SSH_H_realspace}) in real space, with the uniform and disorder-free hopping constants $t_1$ and $t_2$. The results are obtained by scanning $t_1$ and for each $t_1$, the input data set is composed of ordered $M=200$ feature vectors from $t_2 \in \left[-2t_1, 2t_1 \right]$ (see the main text for details). There are in total three topological sectors detected, as indicated by the degree of degeneracy of the largest eigenvalue $\lambda_k^t$ in (b). (c) Images of the data set in the diffusion space. The color code indicates different values of $t_2/t_1$. The $k$-means clustering method (with $k=3$) is used in (c) to automatically retrieve the topological transitions in (a). (b) and (c) are plotted with a representative value of $t_1=1$. Other parameters used are $L=2N_l=80$, $\epsilon=0.03$, and $t=50$.}
\label{fig:clean_SSH_realspace}
\end{figure}

Based on the mathematical fact~\cite{Cowling2019} that the Fourier transform of a $\mathbb{L}^1\left(\mathbb{R}\right)$ [$\mathbb{L}^2\left(\mathbb{R}\right)$] space is the $\mathbb{L}^{\infty} \left(\mathbb{R}\right)$ [$\mathbb{L}^2\left(\mathbb{R}\right)$] space, the corresponding dual distance metrics in real space can be given by the $\mathbb{L}^1$-norm (dual to the CD) and the $\mathbb{L}^2$-norm (dual to the ED), respectively. In the following, we discuss the unsupervised learning of the tight-binding SSH model and a locally disordered SSH model in real space, respectively. Our results show that the performance of the $\mathbb{L}^2$-norm distance (real-space ED) is also suboptimal compared to that of the $\mathbb{L}^1$-norm distance (real-space dual CD)~\cite{Note_Discussion_Ren}.

The clustering of different Hamiltonians or density matrices in real space, in an unsupervised manner, is as interesting as significant. In real space, the calculation of topological invariants is not as straightforward as in momentum space, where topological phases are distinguished by different homotopy classes of maps from the BZ to the Hamiltonian space.

Below we present clustering results for a real-space SSH chain with $N_l$ number of unit cells (i.e., $L=2N_l$ is the number of lattice sites). The Hamiltonian is 
\begin{equation}
\label{eq:SSH_H_realspace}
H = \sum_{n=1}^{N_l} a_n^{\dagger} \hat{h}_n a_{n+1} + h.c. + a_n^{\dagger} \hat{\mu}_n a_{n},
\end{equation}
where 
\begin{equation}
\hat{h}_n = \frac{t_{n2}}{2}(\sigma_x-i\sigma_y), \ \ \hat{\mu}_n = t_{n1} \sigma_x, 
\end{equation}
and $a_n^{\dagger} =\left( c_{n1}^{\dagger}, c_{n2}^{\dagger}\right)$, with $ c_{ni}^{\dagger}$ being the fermion creation operator at the $i$-th site in the $n$-th unit cell. The intra- and inter-cell hopping constants are given by $t_{n1}$ and $t_{n2}$, respectively. The hermitial conjugates of the creation operators give the corresponding fermion annihilation operators. Note that the hopping constants can be disordered, in which case the quasi-momentum is not well-defined. We assume periodic boundary conditions.

As in many machine learning tasks for physics, where the algorithm is fed with samples from numerical simulations or experimental data, here we first solve the energy spectrum and the corresponding eigenvectors of the Hamiltonian in Eq.~(\ref{eq:SSH_H_realspace}), and then take the density matrix within the half lower bands as the input data for the clustering algorithm. 

This is analogous to feeding the algorithm with the lower band states in the two-band model in momentum space. Specifically, the input data is given by the (unnormalized) denstiy matrix $\rho = \sum_n^{N_l} \rho_n$ in the position representation~\cite{LongPRL2020}, where the $\rho_n = \left| \psi_n \right \rangle \left \langle \psi_n \right|$ are the density matrices of the $n$-th energy band $E_n$ (from bottom to top). Then, the diffusion map algorithm and the matrix $\mathbb{L}^1$-norm distance (dual to the CD in momentum space) are used for the unsupervised learning. The similarity matrix from the $\mathbb{L}^p$-norm is given by 
\begin{equation}
{\cal{K}}^p_{ij} = \mathrm{exp}\left(- \ \frac{\lVert\mathbf{x}_i - \mathbf{x}_j \rVert^2_{\mathbb{L}^p}}{2\epsilon L^2}\right),
\end{equation}
where $\mathbf{x}_i$ is the $i$-th density matrix in the vector form.
\begin{figure}[t]
\centerline{\includegraphics[height=5in,width=3.3in,clip]{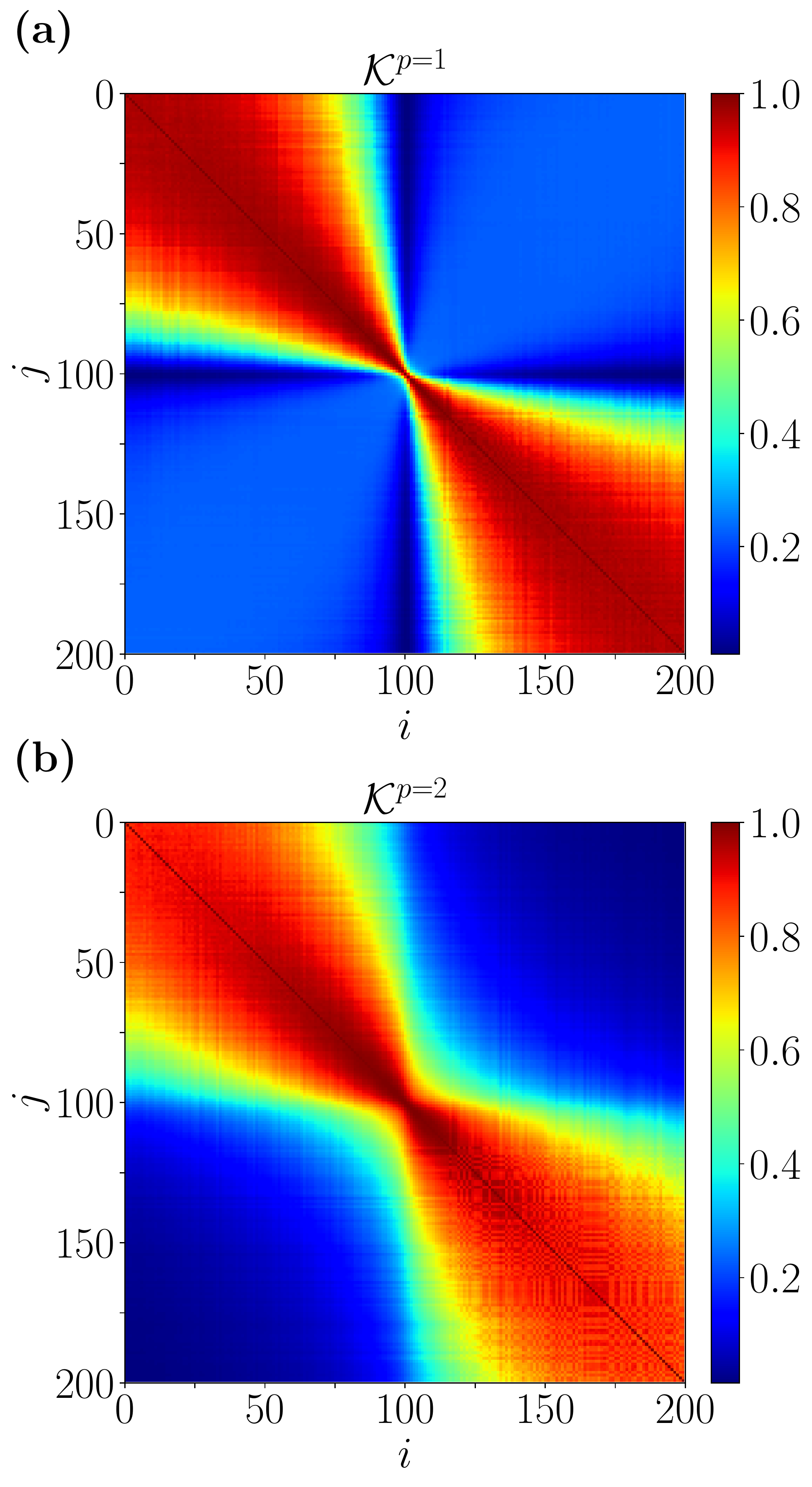}}
\caption{(Color online) Similarity matrix constructed from samples of the locally disordered SSH model in real space. Compared are ${\cal{K}}^p$ with (a) the $\mathbb{L}^1$-norm distance (real-space dual Chebyshev distance) and (b) the $\mathbb{L}^2$-norm distance (real-space Euclidean distance). $M=200$ input data points are uniformly sampled in an \emph{ordered} manner within the range $w \in [-1, 1]$. The middle positions with $i, j = 100$ correspond to $w=0$, where a clear transition and phase boundary can be seen in (a). We have used $L=2N_l=400$ lattice sites. The resolution parameters are $\epsilon = 0.33$ in (a) and $\epsilon = 1.0 \times 10^{-4}$ in (b), which are obtained by minimizing the respective mean squared errors with respect to the ideal similarity matrix, where matrix elements for intra-cluster data points equal one and equal zero for the others. Compared to (a), (b) takes a smaller of several orders resolution parameter $\epsilon$. The dual Chebyshev distance metric in (a) sharpens the feature of the topological transition.}
\label{fig:Dis_SSH_Kmatrix}
\end{figure}

First we consider uniform and disorder-free hopping constants: $t_{n1}=t_1 > 0$ and $t_{n2}=t_2$. 
Figure~\ref{fig:clean_SSH_realspace}(a) shows the learned transition lines in the $t_1-t_2$ phase diagram, with the $\mathbb{L}^1$-norm kernal matrix. There are three clusters [as indicated in Fig.~\ref{fig:clean_SSH_realspace}(b)] and two critical points detected in total.  Plotted in Fig.~\ref{fig:clean_SSH_realspace}(c) are the images of the input data set in the two-dimensional diffusion-mapped space. Detailed hyperparameters are provided in the caption of Fig.~\ref{fig:clean_SSH_realspace}.

Secondly we consider the situation with locally disordered hopping constants~\cite{LongPRL2020}, with $t_{n1} = 1-w r_{1n}$ and $t_{n2} = 1+w r_{2n}$, where $r_{n1}$ and $r_{2n}$ are locally independent random numbers taken from uniform distributions within $(0, 1)$, and $w \in [-1, 1]$ is the disorder strength. The locally random hopping energies break the spatial translation symmetry of the system and the momentum-space description will be no longer valid. As a consequence, Brillouin zone-based calculations of topological invariants are not possible anymore, and hence cannot be used as shortcuts to discover topological transitions. 

Our implementation of the manifold learning with diffusion map and the kernel built from the $\mathbb{L}^1$-norm distance between density matrices lead to a critical line at $w=0$ for this disordered model. Use of the $\mathbb{L}^2$-norm, on the other hand, does not deliver this (correct) result. 

We plot in Fig.~\ref{fig:Dis_SSH_Kmatrix} the comparison between the kernels constructed from the $\mathbb{L}^1$-norm and the $\mathbb{L}^2$-norm distances, respectively, for this disordered model. $M=200$ data points are uniformly sampled within $w \in [-1, 1]$ in an ordered fashion, where $i, j = 100$ corresponds to $w=0$, the critical point. Figure~\ref{fig:Dis_SSH_Kmatrix}(a) shows two nearly ideal clusters separated by $w=0$, outperforming the result shown in Fig.~\ref{fig:Dis_SSH_Kmatrix}(b) given by the $\mathbb{L}^2$-norm distance. In addition, compared to the $\mathbb{L}^1$-norm kernel, the $\mathbb{L}^2$-norm kernel in real space requires a much smaller resolution parameter $\epsilon$, and the application of the diffusion map does not deliver the correct clustering.

\section{Discussion}
Our manifold learning explores and includes the (topologically) characteristic distances between data points. It thus can also cluster samples consisting of energy spectra or eigenvectors, sampled from numerical simulations or extracted from experimental data. Therefore, it can be applied to cluster topological phases in new problems and real materials. The advantage of this method in learning the topological quantum phase transitions is its good interpretability in momentum and (dual) real spaces. With its successful interpretation demonstrated in above benchmark models, it paves the way towards learning more complicated materials, in an unsupervised or semi-supervised manner.

Conventional theoretical approaches studying topological phase transitions are usually formula driven, where
a mathematical expression for the topological invariant is calculated, and the transition of this quantity at
certain critical values of model parameters indicates the occurrence of topological quantum phase transitions. Indeed, to precisely determine the phase boundaries, the numbers of samples required in the two approaches are the same. However, expressions for topological invariants are in general based on prior knowledge, such as the presence of symmetries in the Hamiltonian, and are thus not easily available, for instance, in real space or in the presence of disorders. 

In comparison, unsupervised machine learning methods are data driven, which automatically explore the intrinsic correlations and patterns in a set (manifold) of Hamiltonians or quantum states, without prior detailed analysis. Manifold learning can be a precursor approach for extracting patterns in a data manifold composed of Hamiltonians or quantum states, and is expected to prove its advantage in scenarios where the formulae for topological invariants are unknown or hard to compute. In practice, combining the two approaches may prove most efficient.

\section{Conclusions}
In summary, we have leveraged the $\mathbb{L}^{\infty}$-norm distance in momentum space, as well as the (dual) $\mathbb{L}^1$-norm distance in real space, as approximative topologically characteristic distance measures, and shown how to embed them in the unsupervised manifold learning to successfully retrieve topological quantum phase transitions. In the benchmark Su-Schrieffer–Heeger model and the Qi-Wu-Zhang model considered, the critical lines in the phase diagrams were precisely identified without pre-training. 

We have shown that, compared to the Euclidean distance, the Chebyshev distance in momentum space and the dual $\mathbb{L}^1$-norm distance in real space sharpen the characteristic features of topological quantum phase transitions, making them easier to be retrieved by machine learning methods. This was inspired by the fidelity-susceptibility indicator for topological quantum phase transitions, as well as the non-Euclidean structure of the data set.

In view of the good interpretability and demonstrated performance on several benchmark models in momentum and real spaces, manifold learning has the potential to widen our understanding of topological features in quantum systems, and may find applications in discovering exotic topological quantum phase transitions in momentum or real space, both in theoretical models and in real materials.

\emph{Note added.} After our work for momentum-space learning was completed, two related preprints~\cite{ScheurerPRL2020,LongPRL2020} appeared, studying unsupervised clustering of topological states with different emphases compared to our work.

\paragraph*{Acknowledgments.} We acknowledge helpful discussions with Yu-Ran Zhang and Zhengyang Zhou.
T.L. acknowledges support from the Grant-in-Aid for a JSPS Foreign Postdoctoral Fellowship (P18023).
F.N. is supported in part by: NTT Research,
Army Research Office (ARO) (Grant No. W911NF-18-1-0358),
Japan Science and Technology Agency (JST) (via the CREST Grant No. JPMJCR1676),
Japan Society for the Promotion of Science (JSPS) (via the KAKENHI Grant No. JP20H00134, 
and the grant JSPS-RFBR Grant No. JPJSBP120194828), and
the Foundational Questions Institute Fund (FQXi) via the Grant No. FQXi-IAF19-06.


\appendix

\section{Similarity matrices for the SSH and quenched SSH models}
Here we compare the similarity matrices built from the ED and the CD, for the SSH and the quenched SSH models, respectively. 
\begin{figure}[t]
\centerline{\includegraphics[height=5in,width=3.3in,clip]{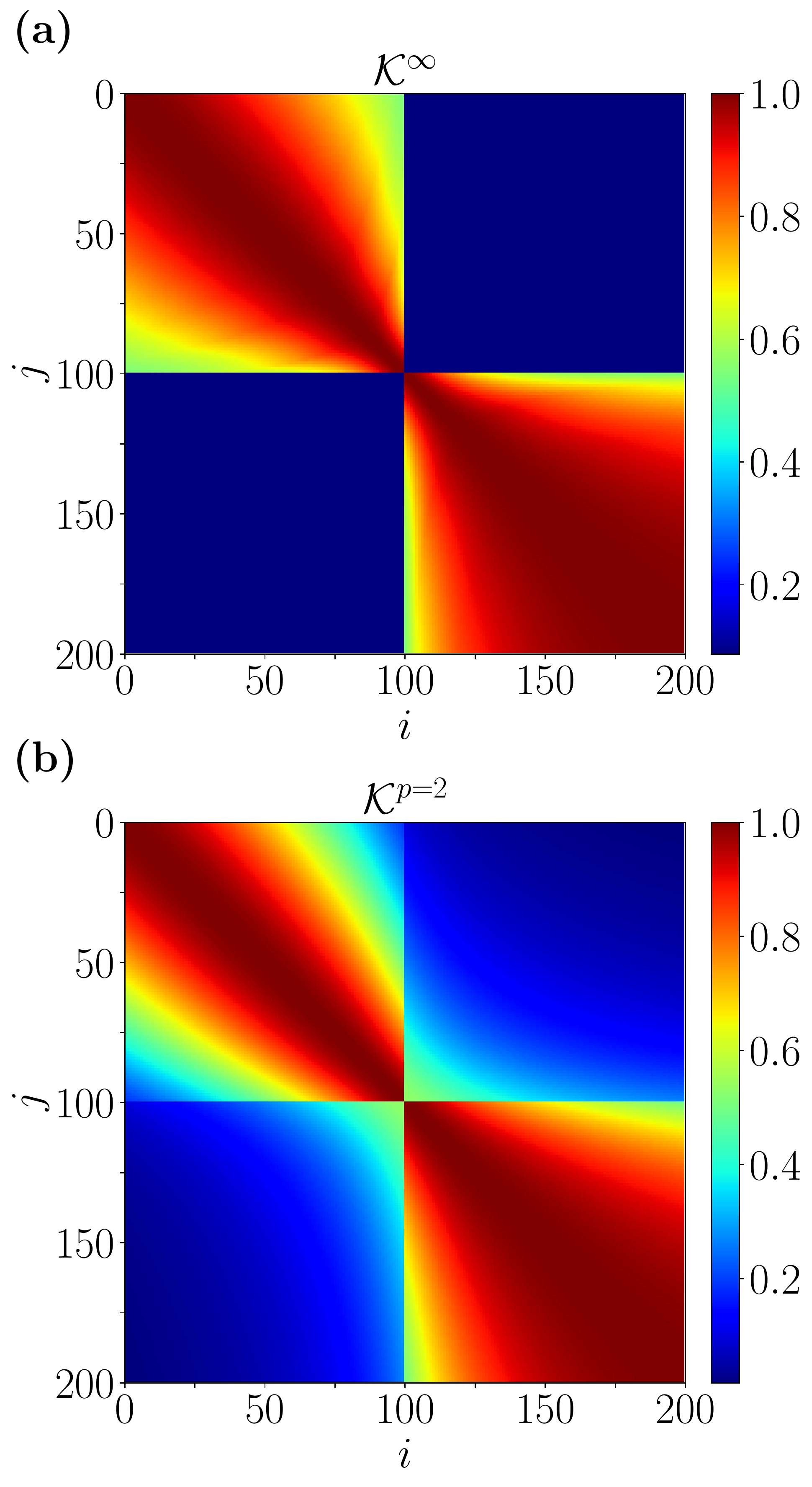}}
\caption{(Color online) Similarity matrix ${\cal{K}}^p$ for the SSH model built from the $\mathbb{L}^p$-norm distance metric: Compared are the (a) Chebyshev distance (CD) with $p=\infty$ and (b) the Euclidean distance (ED) with $p=2$. $M=200$ input data points (reshaped unit Bloch vectors) are uniformly sampled in an \emph{ordered} manner within the range $t_2 \in (0, 2t_1]$, with $t_1 = 1$. The one-dimensional Brillouin zone is sliced into $N=32$ patches. Compared to (b), (a) shows two nearly ideal clusters with good 
(poor) intra- (inter-) cluster connectivity (corresponding two squares in the diagonal direction). The resolution parameters are $\epsilon = 0.41$ in (a) and $\epsilon = 2.68 \times 10^{-6}$ in (b), which are obtained by minimizing the respective mean squared errors with respect to the ideal similarity matrix, where matrix elements for intra-cluster data points equal one and equal zero for the others.}
\label{fig:K_matrix_SSH}
\end{figure}
\begin{figure}[t]
\centerline{\includegraphics[height=5in,width=3.3in,clip]{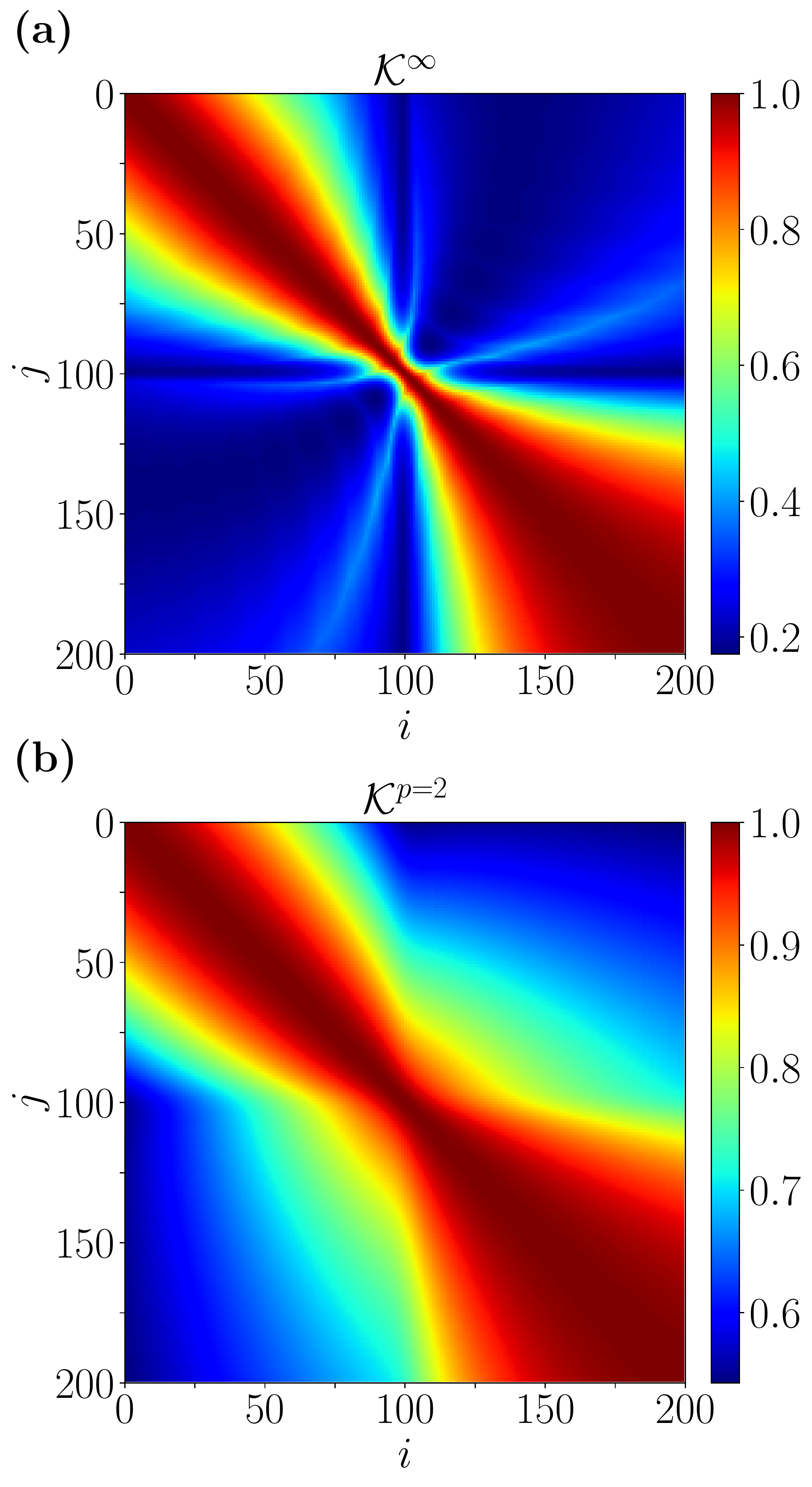}}
\caption{(Color online) Similarity matrix ${\cal{K}}^p$ of the quenched SSH model built from the $\mathbb{L}^p$-norm distance metric: Compared are the (a) Chebyshev distance (CD) with $p=\infty$ and (b) the Euclidean distance (ED) with $p=2$. $M=200$ input data points (reshaped unit Bloch vectors) are uniformly sampled in an \emph{ordered} manner within the range $J_1 \in (0, 2J_0]$, with $J_0 = 1$. The $(t, k)$-space Brillouin zone is sliced into $32 \times 32$ patches. Compared to (b), (a) shows two nearly ideal clusters with good (poor) intra- (inter-) cluster connectivity (corresponding two squares in the diagonal direction). The resolution parameters are $\epsilon = 0.57$ in (a) and $\epsilon = 2.9 \times 10^{-4}$ in (b), which are obtained by minimizing the respective mean squared errors with respect to the ideal similarity matrix, where matrix elements for intra-cluster data points equal one and equal zero for the others.}
\label{fig:K_matrix_qSSH}
\end{figure}
In both cases, $M=200$ data points (i.e., reshaped unit Bloch vectors) are sampled in an ordered manner, and each cluster has equal number of samples (i.e., $M/2$ for even $M$). Then the ideal similarity matrix is given by 
\begin{equation}
{\cal{K}}^{\mathrm{ideal}} = {\cal{I}}_{M/2} \oplus {\cal{I}}_{M/2},
\end{equation}
where ${\cal{I}}_k$ is the $k \times k$ matrix with all matrix entries being $1$. For each case, the value of the resolution parameter $\epsilon$ is obtained by minimizing the mean squared error with respect to this ideal similarity matrix, i.e., 
\begin{equation}
\mathop{\mathrm{min}}\limits_{\epsilon} \frac{1}{M^2} \sum_{ij} \left({\cal{K}}^p_{ij} - {\cal{K}}^{\mathrm{ideal}}_{ij} \right)^2.
\end{equation}

Shown in Fig.~\ref{fig:K_matrix_SSH}(a) and Fig.~\ref{fig:K_matrix_SSH}(b) are the similarity matrices 
for the SSH model, built from the CD and the ED, respectively, where the sampling parameter range is $t_2 \in (0, 2 t_1]$ with $t_1 =1$. Other parameters can be found in the caption of Fig.~\ref{fig:K_matrix_SSH}. 

Shown in Fig.~\ref{fig:K_matrix_qSSH}(a) and Fig.~\ref{fig:K_matrix_qSSH}(b) are the similarity matrices 
for the quenched SSH model, built from the CD and the ED, respectively. The sampling parameter range is $J_1 \in (0, 2 J_0]$ with $J_0 =1$. Also, other parameters can be found in the caption of Fig.~\ref{fig:K_matrix_qSSH}.

\section{Learning the Qi-Wu-Zhang model with the Euclidean distance}
\label{sec:append_QWZ_ED}
\begin{figure}[t]
\centerline{\includegraphics[height=3.2in,width=3.5in,clip]{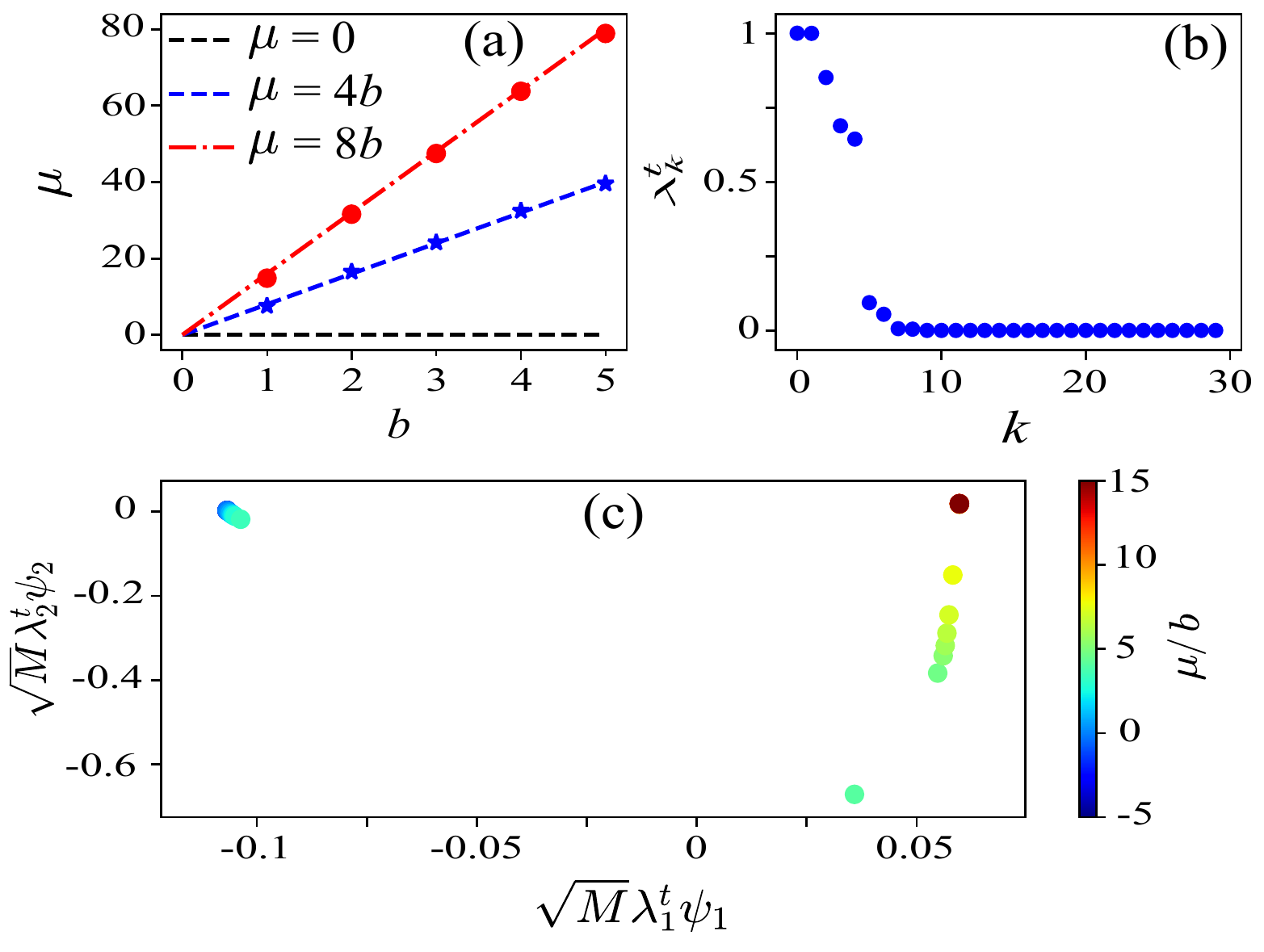}}
\caption{(a) Unsupervised learning of the Qi-Wu-Zhang model using the diffusion map algorithm with the Euclidean distance. All the parameters used are the same as that in Fig.~\ref{fig:QWZ_diff_map} in the main text, except for a carefully fine-tuned value of the resolution parameter $\epsilon=2 \times 10^{-5}$ for $b=1$ and $\epsilon=6 \times 10^{-6}$ for the other values of $b$, and $t=100$. This is the best result one can obtain by fine-tuning the value of 
$\epsilon$. In (a), only two critical lines (the red dots and blue stars) can be retrieved, while the first two clusters in the correct phase diagram (separated by $\mu=0$) are mixed into one cluster in this case. One cannot read out the correct number of clusters from (b), either. In (c), the color code indicates different values of $\mu/b$. (b) and (c) are plotted with a representative value of $b=1$.}
\label{fig:QWZ_ED} 
\end{figure}
Here we provide the learning results for the QWZ model in the main text, with the diffusion map, and using the ED as a distance metric. The ED in general gives incorrect results. By carefully fine-tuning the resolution parameter $\epsilon$, the best result for the ED is given in Fig.~\ref{fig:QWZ_ED}. The carefully fine-tuned resolution parameter $\epsilon$ is much smaller than that used in the CD case, indicating that the ED is not a good distance metric discriminating different clusters in the data set. Only two critical lines (the red dots and blue stars) can be retrieved, while the first two clusters in the correct phase diagram (separated by $\mu=0$) are mixed into one cluster in this case.

\section{Unsupervised learning with principal component analysis and Isomap for the Qi-Wu-Zhang model}
\label{sec:append_QWZ_Isomap}
To further clarify the suboptimal performance of linear models such as the principal component analysis (PCA), and the advantage of the Chebyshev distance combined with manifold learning, here we present clustering results for the QWZ model from the PCA and give a comparison between the Isomap using the CD and ED, respectively. We use the QWZ model because it exhibits multiple topological transitions and can thus be used to test various methods.
\begin{figure}[t]
\centerline{\includegraphics[height=4.2in,width=3in,clip]{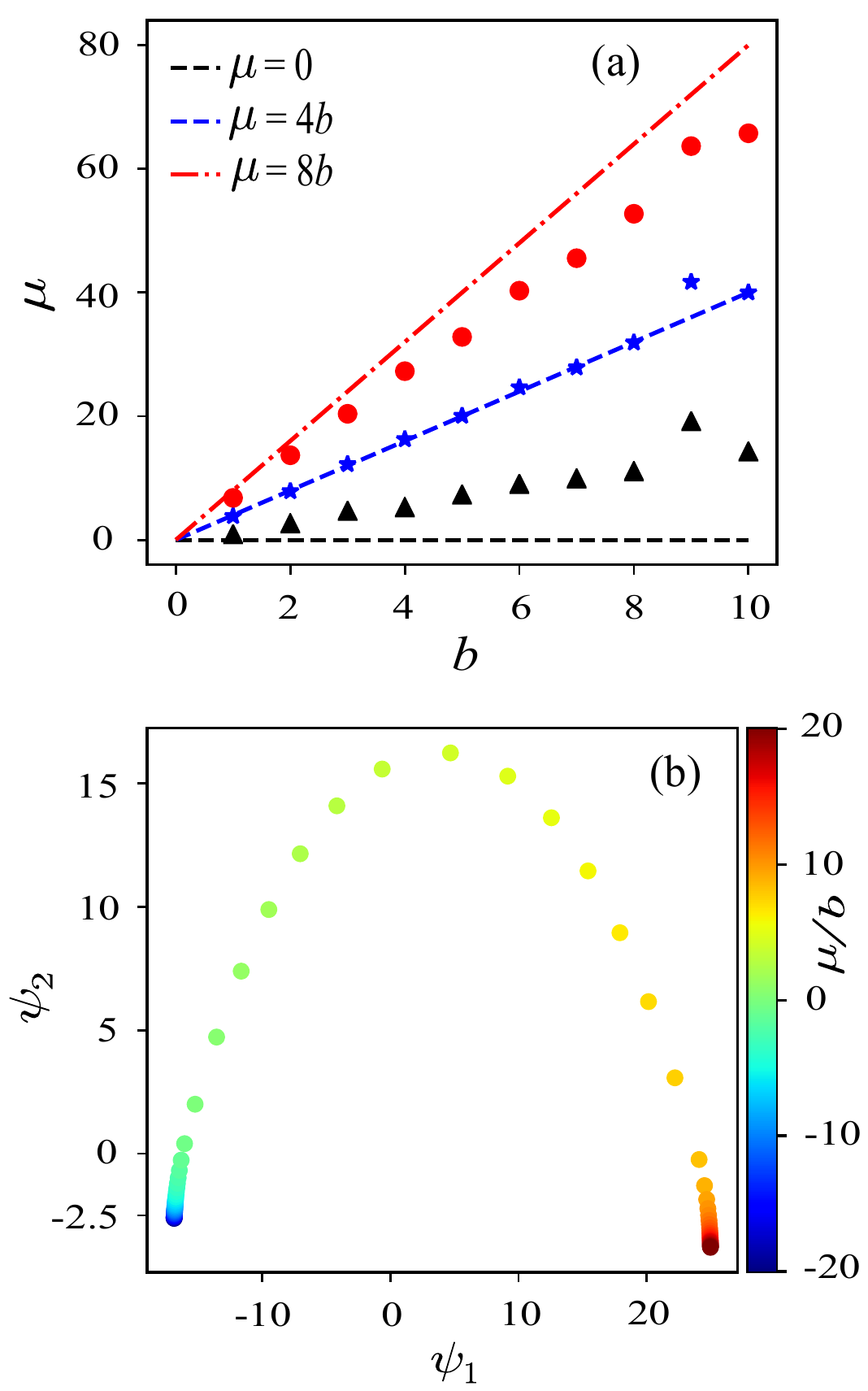}}
\caption{(Color online) Unsupervised learning of the Qi-Wu-Zhang model using the principal component analysis, with (a) the retrieved critical lines (dark triangles, blue stars and red dots) and (b) data images in the first two principal components $\psi_1$ and $\psi_2$. The data set consists of uniformly sampled unit Bloch vectors as in the diffusion map algorithm in the main text, with the chemical potential varying from $\mu= -20b$ to $\mu= 20b$. The $k$-means clustering method (with $k=4$) is then used in (b) to automatically retrieve the topological transition in (a). The hopping energy in (b) is set to be a representative value $b=1$. $M=1000$ data points are sampled and $N=20$ is used.}
\label{fig:PCA_QWZ}
\end{figure}

PCA projects data points along several most dispersed directions, i.e., directions with large variances. It is a linear unsupervised learning model and cannot uncover nonlinear structures in the data set. Note that the success of PCA in clustering certain topological phases~\cite{YeHuaNatPhys2017} is based on preprocessing the raw data (the Hamiltonian or the wave function), where a nonlinear transformation will be performed in the first step. For instance, Ref.~\onlinecite{YeHuaNatPhys2017} took the entanglement spectrum of the Kitaev model as the input of the PCA algorithm. This makes sense because the topological features and transitions can be directly read from the entanglement spectrum~\cite{HaldanePRL2008,PollmannPRB2010}. In contrast, for many applications where the input contains raw wave functions or Hamiltonians, linear learning models are not optimal choices.
\begin{figure}[t]
\centerline{\includegraphics[height=4in,width=3.5in,clip]{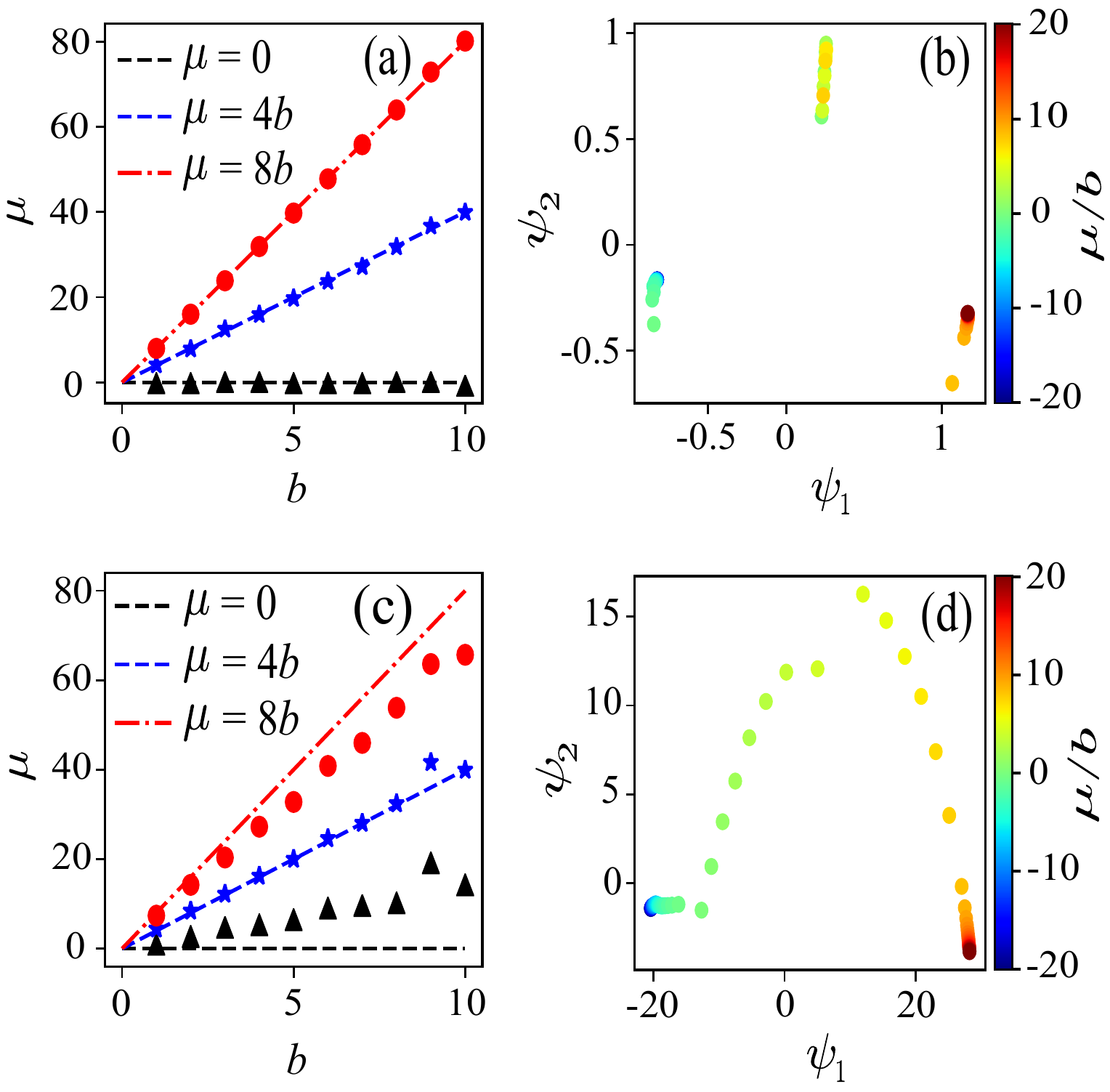}}
\caption{(Color online) Unsupervised manifold learning of the Qi-Wu-Zhang model using Isomap. (a)-(b) are obtained using the Chebyshev distance to construct nearest-neighbor graph in the isometric embedding. (c)-(d) are obtained using the Euclidean distance. In (a) and (c), the dark triangles, blue stars and red dots are the learned critical points, respectively. (b) and (d) are the respective images of sampled data points in the low-dimensional embedded space. In both cases, the data set consists of uniformly sampled unit Bloch vectors as in the diffusion map algorithm in the main text, with the chemical potential varying from $\mu= -20b$ to $\mu= 20b$. The $k$-means clustering method (with $k=4$) is used to automatically retrieve the topological transitions from the embedded space. In both (b) and (d), we take $b=1$ for a representative value. Other hyperparameters used are $M=1000$, $N=20$ and the number of neighbors used in the Isomap algorithm is $500$.}
\label{fig:Isomap_QWZ}
\end{figure}

In Fig.~\ref{fig:PCA_QWZ} we show the learning results for the QWZ model given by the PCA. It fails to retrieve the correct topological transition lines in Fig.~\ref{fig:PCA_QWZ}(a), and the data points projected to the first two principal components do not exhibit cluster structures that are linearly separable.

In comparison to PCA, Isomap~\cite{TenenbaumScience2000} is one of the important manifold learning methods that achieves nonlinear dimensionality reduction in an unsupervised manner. The manifold geodesic distance, approximated from the shortest distance on the neighborhood graph constructed from the data set, should then be used to characterize the similarities. The following step is to embed the data points into a meaningful low-dimensional Euclidean space based on this adaptive distance, where a conventional clustering method (e.g., $k$-means) can be used to detect the structure of the data set. There is another degree of freedom in this process, i.e., the original distance metric used in constructing the neighborhood graph. 

In Fig.~\ref{fig:Isomap_QWZ}, we present a comparison for the Isomap learning of the QWZ model with the CD and the ED, respectively, to construct the neighborhood graph. Figure~\ref{fig:Isomap_QWZ}(a)-(b) are obtained using the CD to construct the nearest-neighbor graph in the isometric embedding. Figure~\ref{fig:Isomap_QWZ}(c)-(d) are obtained using the ED. One can read from Fig.~\ref{fig:Isomap_QWZ}(a),(c) that the CD performs better than the ED, where the dark triangles, blue stars and red dots are the learned critical points, respectively. Figure~\ref{fig:Isomap_QWZ}(b),(d) are the respective images of sampled data points in the low-dimensional embedded Euclidean spaces.

\section{Learning over wave functions}
\label{sec:append_wavefunction}
\begin{figure}[t]
\centerline{\includegraphics[height=5in,width=3.3in,clip]{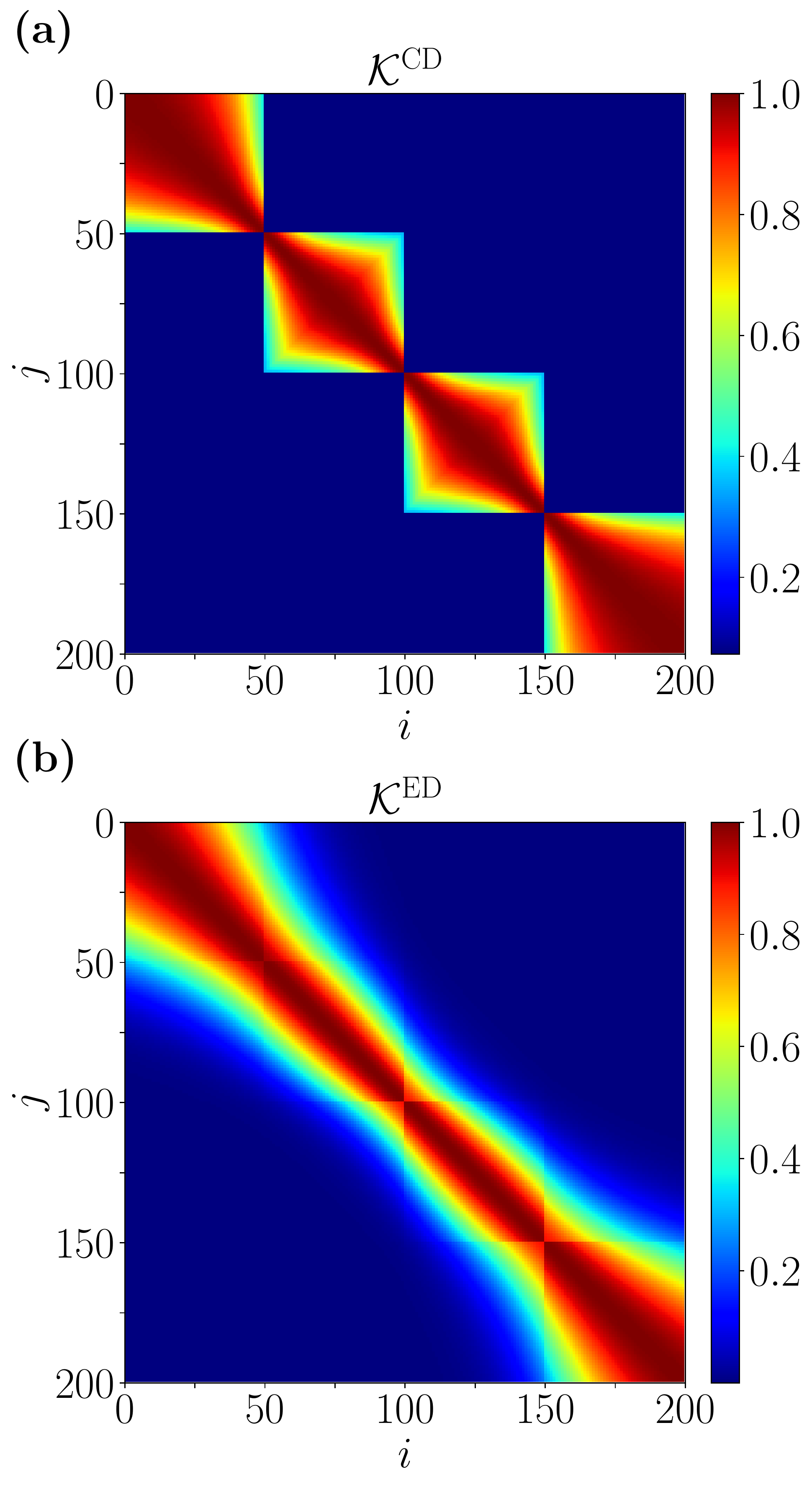}}
\caption{(Color online) Similarity matrix with ground state wave functions of the Qi-Wu-Zhang model as input. Compared are (a) Chebyshev distance (CD) and (b) the Euclidean distance (ED). $M=200$ input data points are uniformly sampled in an \emph{ordered} manner within the range $\mu \in [-4b, 12b]$, with $b = 0.2$. The $2$D Brillouin zone is sliced into $32 \times 32$ patches. The resolution parameters are $\epsilon = 0.382$ in (a) and $\epsilon = 0.0326$ in (b), which are obtained by minimizing the respective mean squared errors with respect to the ideal similarity matrix, where matrix elements for intra-cluster data points equal one and equal zero for the others.}
\label{fig:WF_Kmatrix_QWZ}
\end{figure}
In many physical scenarios, it is useful to use wave functions as the input data and detect topological transitions of the ground state. Then the similarity measure used for supervised or unsupervised learning can be defined through the local fidelity of quantum states 
\begin{equation}
F^{ij}_k = \left| \langle \psi^i_k | \psi^j_k \rangle \right|^2,
\end{equation}
where $k$ is the local index, and $i$ and $j$ are the sample indices, respectively. The squared ED corresponds to an averaged (pseudo) distance measure over the local index~\cite{ScheurerPRL2020}: 
\begin{equation}
d^2_{\mathrm{E}} (i, j) = \frac{1}{{\cal{N}}} \sum_k \left( 1 - F^{ij}_k \right),
\end{equation}
with ${\cal{N}}$ the dimension of the wave-function feature vector. The squared CD is
\begin{equation}
d^2_{\mathrm{C}} (i, j) = \mathop{\mathrm{max}}\limits_k \left( 1 - F^{ij}_k \right).
\end{equation}
For quantum states in momentum space, the local index is the momentum $\textbf{k}$. Here the (pseudo) distances are effective similarity measures, which may not satisfy the triangle inequality required by the mathematical definition of distance metrics.

The corresponding similarity (kernel) matrices are:
\begin{equation}
\label{eq:wave_function_Kmatrix}
{\cal{K}}^{\mathrm{CD (ED)}}_{ij} = \mathrm{exp}\left(- \ \frac{d^2_{\mathrm{C (E)}} (i, j)}{\epsilon}\right).
\end{equation}

The Euclidean distance takes the average over local EDs at each momentum in the Brillouin zone (BZ), and this average process smears out the characteristics of topological transitions, where only at certain special points in the BZ (not at all points) sign changes may occur and the local wave functions abruptly become orthogonal across the phase transition. 

Note that this is different from the divergence of the averaged fidelity susceptibility (over the BZ) as an indicator of topological quantum phase transitions (see, e.g., Ref.~\onlinecite{MaPRB2010}). The averaged fidelity susceptibility can be strongly peaked at the transition point if some local ones become nearly divergent. The use of the Chebyshev distance instead of the ED was somewhat inspired from this. We have verified that the unsupervised learning over wave functions gives the same result as that over the Bloch vectors for the two band models, e.g., the QWZ and SSH models. 

In Fig.~\ref{fig:WF_Kmatrix_QWZ} we show the similarity matrix built from the wave functions for the QWZ model, with the CD and the ED in the kernel function, respectively. Their structures are similar to Fig.~\ref{fig:K_matrix} in the main text. For more generic and complex models in momentum space, such as symmetry-protected topological orders, multi-band models, and Chern insulators, the topological critical points can be retrieved from learning over proper wave functions, which are maps from the BZ, as feature descriptors, and by using the CD combined with manifold learning algorithms.


\begin{thebibliography}{91}
\bibitem [{Par()}]{ParameterRangeNote}%
  \BibitemOpen
  \href@noop {} {}\bibinfo {note} {Further extension of this parameter range
  does not change the result. But more samples will be required when $b$ is
  large, to precisely retrieve the critical points.}\BibitemShut {Stop}%
\end{thebibliography}
%

\end{document}